\documentclass[showpacs,preprintnumbers,amsmath,amssymb]{revtex4}

\usepackage{graphicx}
\usepackage{dcolumn}
\usepackage{bm}
\usepackage{epsfig}
\usepackage{amsfonts}

\begin{document}

\title{Probability densities for the sums of iterates of the sine-circle map 
in the vicinity of the quasi-periodic edge of chaos}

\author{Ozgur Afsar$^{1,}$}
\email{ozgur.afsar@ege.edu.tr}
\author{Ugur Tirnakli$^{1,2,}$}
\email{ugur.tirnakli@ege.edu.tr}
\affiliation{
 $^1$Department of Physics, Faculty of Science, Ege University, 35100 Izmir, Turkey\\
$^2$Division of Statistical Mechanics and Complexity, Institute of Theoretical and Applied Physics (ITAP) 
Kaygiseki Mevkii, 48740 Turunc, Mugla, Turkey
}

\date{\today}

\begin{abstract}
We investigate the probability density of rescaled sum of iterates of
sine-circle map within quasi-periodic route to chaos. When the dynamical 
system is strongly mixing (i.e., ergodic), standard Central Limit Theorem 
(CLT) is expected to be valid, but at the edge of chaos where iterates 
have strong correlations, the standard CLT is not necessarily to be valid anymore. 
We discuss here the main characteristics of the probability densities for the 
sums of iterates of deterministic dynamical systems which exhibit quasi-periodic route to chaos. 
At the golden-mean onset of chaos for the sine-circle map, we numerically 
verify that the probability density appears to converge to a $q$-Gaussian with $q<1$ 
as the golden mean value is approached. 
\end{abstract}

\pacs{05.20.-y, 05.45.Ac, 05.45.Pq}

\maketitle

\section{Introduction}
The standard Central Limit Theorem (CLT) asserts that, under appropriate conditions,
the probability density of the sum of large number of independent
identically distributed ({\it iid}) random variables will be approximately normal
(or Gaussian) \cite{cramer}. The first complete proof was given in 1901 by 
Liapunov \cite{liapunov}, who worked analitically with characteristic functions. 
Lindeberg gave a complete proof of the CLT under more general conditions than Liapunov 
in his 1922 paper \cite{lindeberg}. 
The CLT explains why many stochastic processes are of relevance in physics,
chemistry, biology, economics, etc. One of the fundamental importance of CLT 
for statistical mechanics is that if a CLT is valid for the driving forces in a 
many-body system, it is easy to proceed to the formalism of statistical mechanics via the
Langevin and Fokker-Planck approaches.

It is also well-known that there are CLTs for the deterministic dynamical systems. 
Although deterministic dynamical systems have deterministic algorithms and therefore 
never be completely independent, if the assumption of {\it iid} is replaced by the 
weaker property that the dynamical system is sufficiently strongly mixing, then 
various CLTs for these systems are also shown to be 
valid \cite{billingsley,kac,beck,roepstroff,mackey}. 
Here, ``sufficiently strongly mixing'' implies ergodicity and just means asymptotic 
statistical independence for large time differences. 
In this case the standard CLT is valid for deterministic dynamical systems. 
If the iterates of deterministic dynamical systems have strong correlations like 
at the edge of chaos where the system is only weakly chaotic, then the standard CLT is 
not valid. In fact, many complex systems in nature, such as financial data \cite{cortines}, 
earthquakes \cite{caruso}, long-range interacting many body systems \cite{pluchino}, 
exhibit strong correlations among appropriate random variables. 
At least for a class of certain strong correlations (referred to as $q$-independence), 
it has been analytically shown that the attractors are $q$-Gaussians leading
to a $q$-genaralization of the standard CLT \cite{umarov,tsallis,queiros,vignat}. 
$q$-Gaussian distributions, defined as, 

\begin{equation}
P(x)= [(1-(1-q)\beta x^{2}]^{1/(1-q)}, 
\label{q-gaussian}
\end{equation}
(where $\beta$ controls the width of the distribution) 
are the distributions that optimize the nonadditive entropy
$S_q$ (defined to be $S_{q} \equiv \left(1- \sum_i p_i^q\right)/ \left(q-1\right)$), 
on which nonextensive statistical mechanics is based \cite{tccmp,tsallisbook}. 
As $q\rightarrow 1$, $q$-Gaussians recover the Gaussian distribution.

In recent years, many dissipative deterministic dynamical systems like cubic map,
logistic map and logaritmic map are numerically investigated in terms of 
their central limit behaviour \cite{tibet,tirbectsal,tsatir,robledo1,tsalruiz}.  
The common property for all of these examples is that they exhibit period doubling 
route to chaos. 
Another dissipative but high dimensional system known as Kuramoto model has also 
been studied within the same context \cite{andrea}.  
Finally, a very recent work has also appeared as a first attempt of a 
similar analysis of conservative maps \cite{silvio}. 
In this work, we focus our attention to another one-dimensional dissipative dynamical 
system, known as sine-circle map \cite{hilborn,schuster}, which differs from the above-mentioned 
maps since it exhibits quasi-periodic route to chaos. 
We are mainly interested in two important questions: 
(i)~what are the typical probability distributions in the vicinity of the quasi-periodic edge of chaos? 
(ii)~what is the relationship between the degree of quasi-periodicity of trajectories 
and the complete shape of appropriate limit distribution?

Let us start by introducing the sine-circle map

\begin{equation}
\theta_{t+1}=\theta_t +\Omega -\frac{K}{2\pi}\sin{(2\pi\theta_t)}
~~~ \mod(1) \;\; ,
\label{circlemap}
\end{equation}
where $0\leq\theta_t<1$ is a point on a circle and parameter
$K$ (with $K>0$) is a measure of the strength of the nonlinearity. 
It describes dynamical systems possessing a natural frequency $\omega_1$ which are 
driven by an external force of frequency $\omega_2$ 
($\Omega =\omega_1/\omega_2$ is the bare winding number or frequence-ratio parameter)
and belongs to the same universality class of the forced Rayleigh-Benard 
convection \cite{jensen}. 
Winding number for this map is defined to be the limit of the ratio

\begin{equation}
W =\lim_{t\rightarrow\infty} \frac{(\theta_{t}-\theta_0)}{t} \;\; ,
\label{winding}
\end{equation}
where $(\theta_{t}-\theta_0)$ is angular distance traveled after $t$ iterations of the 
map function. 
The map is monotonic and invertible (nonmonotonic and noninvertable) for $K<1$ ($K>1$) and 
develops a cubic inflexion point at $\theta=0$ for $K=1$. 
The winding number $W(\Omega)$ can be numerically computed from Eq.~(\ref{winding}) 
forming a ``Devil's Staircase'' shown in Fig.~1. 
Once mode is locked, it is noted that $W$ does not change for a substantial range of $\Omega$. 
If $\Omega$ belongs to a constant interval of the Devil's staircase 
($\omega_1/\omega_2$ is commensurate), then $W$ is rational number (is also called mod-locked) 
and the behavior of the system is periodic, otherwise, i.e., if $\Omega$ does not belong to 
a constant interval ($\omega_1/\omega_2$ is incommensurate), then $W$ is irrational number 
and the behavior of the system is quasi-periodic.

At the onset of chaos where $K=1$, a set of zero measure and universal scaling dynamics is produced 
at special irrational dressed winding numbers. They can be approximated by a sequence of truncated 
continued fractions. 
The most interesting and well-studied case is the sequence of rational approximants to 
$W_{GM}=(\sqrt{5}-1)/2$ which is called the golden mean and this has the form of an infinite 
continued-fraction 

\begin{equation}
W=\frac{1}{1+\frac{1}{1+\frac{1}{1+...}}} \;\; .
\end{equation}
If fraction lines stop at $n$ denominator, it is the $n$th order approximation to the golden mean. 
It is also easy to see that the $n$th approximation to the golden mean is given 
by $W_n=({F_n}/{F_{n+1}})$, where $F_n$  is $n$th Fibonacci number and 
$W_n$ is the $n$th convergent for the golden mean. 
This states that the sequence of rational numbers $W_n$ converges to the irrational number 
$W_{GM}$ as ${n\rightarrow\infty}$, yielding the frequency-ratio parameter to approach its 
limiting value $\Omega_{\infty}$.

This map has already been studied in \cite{circle1} at the edge of chaos, where the sensitivity to 
initial condition behavior has numerically analyzed for the first time and its connection with the 
nonextensive statistical mechanics has been established. Then, the numerical results given 
in \cite{circle1} have been analytically proven by Hernandez-Saldana and Robledo \cite{robledo}. 
In the remainder of this work, we focus on the probability density of rescaled sum of iterates of 
the sine-circle map in the strongly chaotic regime and at the quasi-periodic edge of chaos 
and discuss the main features of the probability distribution as the quasi-periodic edge of 
chaos is approached.

\begin{figure}
\includegraphics*[height=8cm]{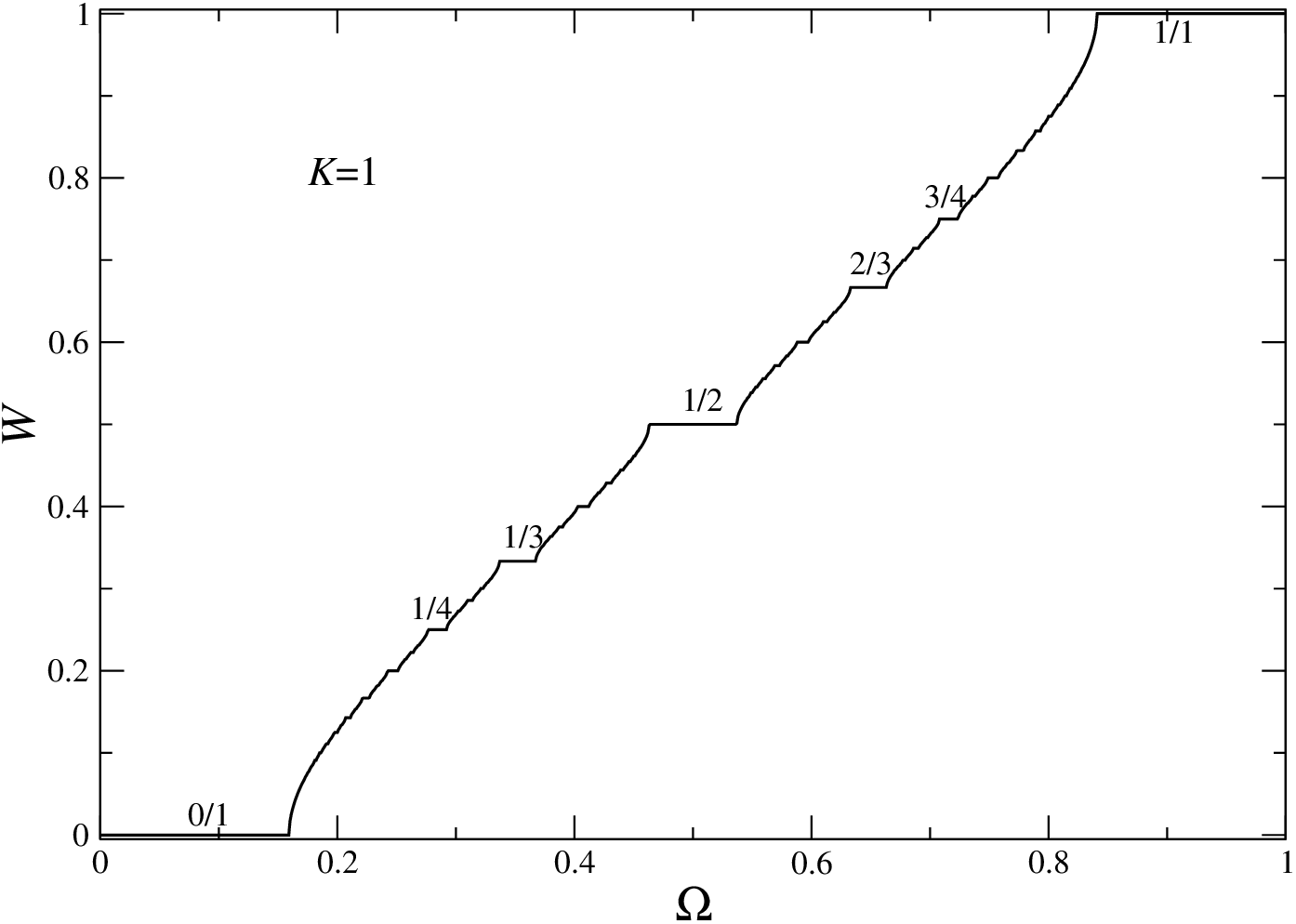}
\caption{Devil's staircase which shows the mode-locking structure of the sine-circle map. 
Some representative rational numbers for which the winding number is locked are given on the curve.}
\end{figure}

\section{Probability densities of the sums of iterates of the sine-circle map}
\subsection{Strongly chaotic case}

Generally, one can consider the sine-circle map given in Eq.~(\ref{circlemap}) 
and investigate the behavior of this dynamical system where it is strongly chaotic. 
When this is the case, implying asymptotic statistical independence, then correlations among 
distributed random variables asymptotically converge to zero, and thus the standard CLT is expected 
to be valid for this case. In order to numerically illustrate this, we can consider the sine-circle map 
at $K=5$ and $\Omega=0.606661063469$ where the map is strongly chaotic and check whether the probability 
distribution of 
 
\begin{equation}
y:= \sum_{i=1}^N \left(\theta_i -\langle \theta \rangle\right) \;\; ,
\label{y}
\end{equation}
becomes Gaussian for $N \to \infty$, after appropriately scaled and centered, regarding the initial value 
as a random variable. Here, the average $\langle ... \rangle$ is calculated as time average. 
The result is given in Fig.~2 where a clear convergence to a Gaussian is evident as expected.

\begin{figure}
\includegraphics*[height=6cm]{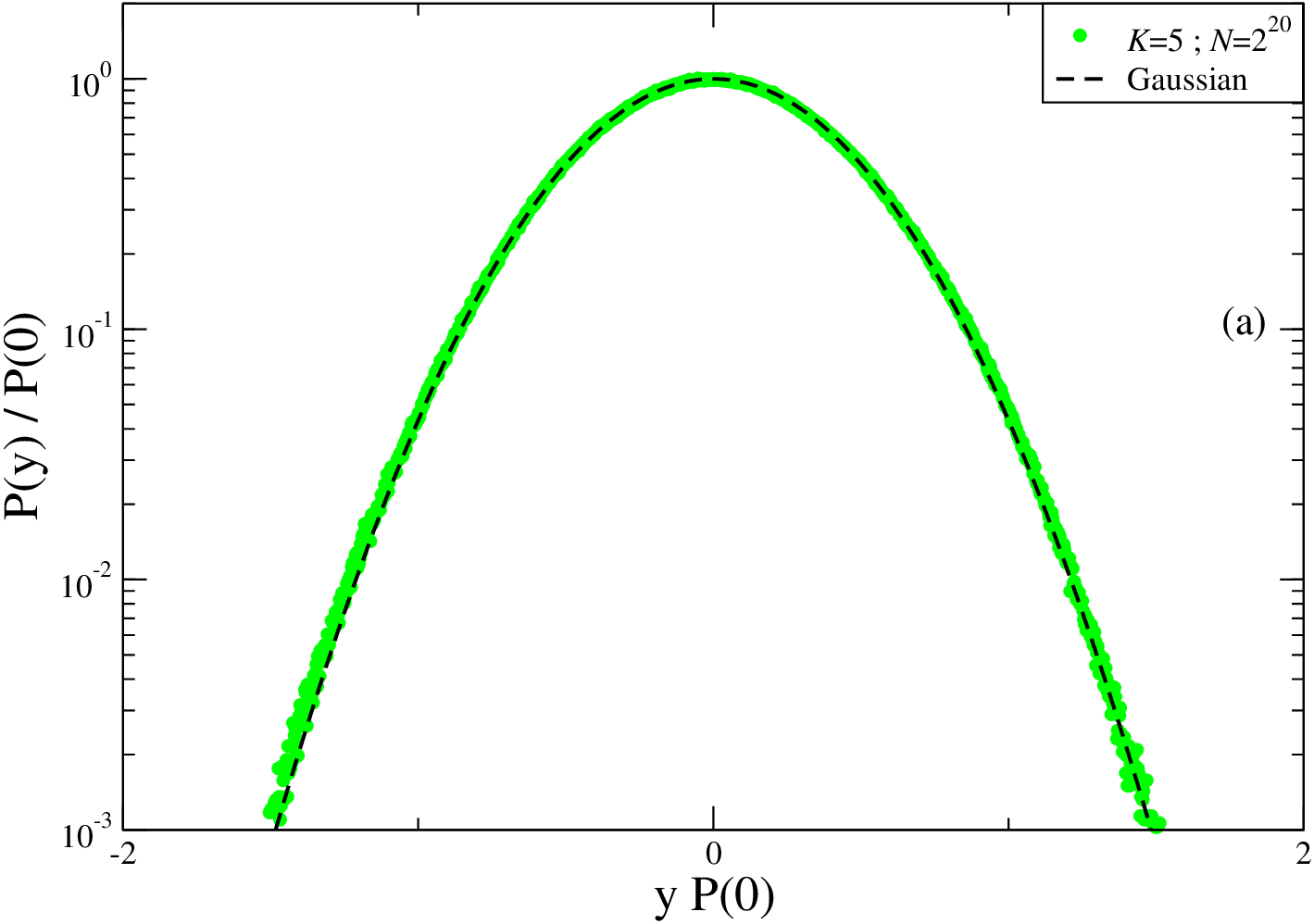}
\includegraphics*[height=6cm]{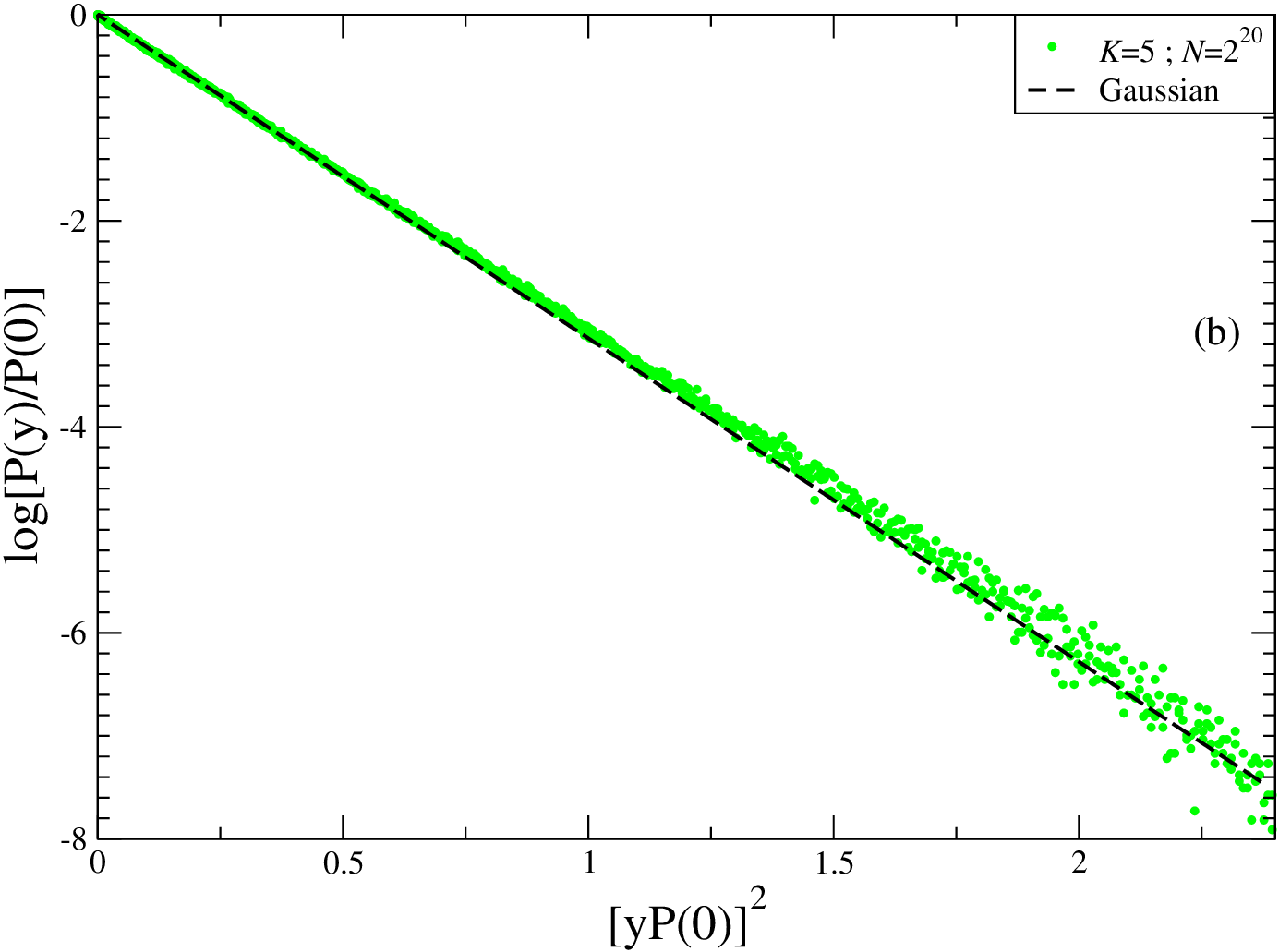}
\caption{(Color online) (a)~Probability distribution function of the sine-circle map for 
$K=5$ and $\Omega=0.606661063469$. Black dashed line corresponds to the Gaussian 
$P(y)=e^{-y^2/(2\sigma^2)}/\sqrt{2\pi\sigma^2}$. (b)~Logarithmic plot of the same data 
where a linear tendency, which corroborates the Gaussian behavior, is appreciated.}
\end{figure}

\subsection{Approaching the quasi-periodic edge of chaos}
We are now at the position to investigate the behavior of the sine-circle map in the vicinity of 
the quasi-periodic edge of chaos for which the standard CLT is not expected to be valid due to 
strong correlations among the iterates of the map. 
At the critical point the system is not ergodic and strongly mixing anymore. 
To numerically calculate the averages $\langle ... \rangle$ in Eq.~(\ref{y}), due to lack of ergodicity, 
it is necessary to take the average over not only a large number of $N$ iterations but also a large 
number of $M$ randomly chosen initial values

\begin{equation}
\langle \theta \rangle = \frac{1}{M} \frac{1}{N} \sum_{j=1}^{M} \sum_{i=1}^N \theta_i^{(j)}\; .
\end{equation}
Although this condition is of no importance at strongly chaotic regime because of ergodicity, 
this is certainly of great importance at the edge of chaos due to non-ergodicity.

At the exact quasi-periodic edge of chaos, we know that $K=1$ and the frequency-ratio parameter 
should be taken as an irrational number, like $\Omega_{\infty}$, corresponding to $W_{GM}$. 
Unfortunately, in a numerical experiment, this parameter can never be obtained with infinite precision 
(which means that the winding number is exactly equal to the golden mean value) but can be approached 
with increasing precision (namely, the winding number is getting closer to the exact golden mean value 
as larger Fibonacci numbers are used). 
In this case, since the golden mean is approached with the ratio of the Fibonacci numbers 
$W_n=(F_n/F_{n+1})$, the trajectory has the period $F_{n+1}$. It means that trajectory 
returns to its starting point on the circle. Therefore, there exist $F_{n+1}$ quasi-periodic bands 
of the attractor within its self-similar structure. 
As a result, the critical behavior of the map can never be numerically studied at the exact chaos 
threshold where the frequency-ratio parameter is expected to be completely irrational corresponding 
to the case of $W$ approaching to the golden mean with infinite precision, but can only be analyzed 
in the vicinity of the critical point, systematically getting closer to it by increasing the finite 
precision of the golden mean adjusting the frequency-ratio parameter. This procedure is exactly the 
same as the one used in \cite{tirbectsal,tsatir} for the same analysis of the logistic map. 
Similar to the case of the logistic map, one can write down a scaling relation \cite{schuster}

\begin{equation}
|\Omega_{n}(K)-\Omega_{\infty}(K)|=\frac{1}{|\delta|^{n+1}} \;\; ,
\label{scaling}
\end{equation} 
where $|\delta|=2.83362...$ is the Feigenbaum constant, $\Omega_{\infty}$ is the value whenever 
$W_{GM}$ is achieved. 
In practice we localize the $\Omega_{n}$ values from the scaling relation (\ref{scaling}) 
and approach the critical point by increasing $n$ values. 
Numerical values used throughout this work have been given in Table~I.

\begin{table}
\caption{\label{tab:Table} Numerical values of $\Omega$ used in the simulations 
and the associated values of $n$, $W_n$ and $N^{*}$.}
\begin{tabular}{|c|c|c|}
\hline \hline 
  $n$                 &    $\Omega_n$       &   $N^{*}=(F_{n+1})^2$  \\ 
\hline \hline
$11$ & $0.606664795117167$ &   $144^2$              \\ \hline
$13$ & $0.606661528216938$ &   $377^2$              \\ \hline
$15$ & $0.606661121349764$ &   $987^2$              \\ \hline
\hline 
$\infty$             & $0.606661063469...$ &  $\infty$             \\ 
\hline \hline
\end{tabular}
\end{table}

As a result of above-mentioned procedure, we are not exactly at the chaos threshold and therefore 
the sum of iterates $\sum_{i=1}^{F_{n+1}} \theta_i$ 
will essentially approach to a fixed value $v =F_{n+1} \langle \theta \rangle$ plus a small correction 
$\Delta v_1$ which describes the small fluctuations of the position of the $(F_{n+1})$th iterate within 
the quasi-periodic band. Therefore, one can write 

\begin{equation}
y_1=\sum_{i=1}^{F_{n+1}} (\theta_i-\langle \theta \rangle ) =\Delta v_1 \;\;, 
\end{equation}
and 

\begin{equation}
y_2=\sum_{i=F_{n+1}+1}^{2F_{n+1}} (\theta_i-\langle \theta \rangle ) =\Delta v_2 \;\;. 
\end{equation}
If we continue this iteration $F_{n+1}$ times respectively, we obtain 
$F_{n+1}$ strongly correlated random variables 
$(y_1,y_2,...,y_{F_{n+1}})=(\Delta v_1,\Delta v_2,...,\Delta v_{F_{n+1}})$ at last. 
Every new fluctuation $\Delta v_j$ is correlated with the others and correlations among these 
random variables decay very slowly if we are close to the irrational frequency-ratio parameter. 
Also, we use the total of these strongly correlated random variables to obtain
appropriate limit distribution function in the vicinity of the quasi-periodic edge of chaos

\begin{equation}
\sum_{j=1}^{F_{n+1}} \Delta v_j=\sum_{i=1}^{(F_{n+1})^2} y_i=\sum_{i=1}^{(F_{n+1})^2} 
(\theta_i -\langle \theta \rangle ) \;\;.
\end{equation}
This clearly implies that, in order to see the convergence to the limit distribution, 
the appropriate number of iterations $N^*$ must be taken as 

\begin{equation}
N^* = (F_{n+1})^2 \;\;\; .
\end{equation}

Our main results are given in Figs.~3 and 4. 
We plot the probability density function in Fig.~3a for $\Omega_n$ values approaching 
$\Omega_{\infty}$ for the first three cases given in Table~I with $n=11$, $n=13$ and 
$n=15$ with appropriate $N^*$ values obtained from the scaling relation.
It is clearly seen from Fig.~3a that the curves have a tendency to converge to a limit 
distribution as the critical point is approached. 
It seems that the limit distribution can be well approximated by a $q$-Gaussian 
with $q=0.925$. 
When $n$ is small (more distant case to the critical point), the data converges to this 
$q$-Gaussian distribution at the more central regions but deviates in the tails. 
As $n$ increases (getting closer to the critical point), this convergence develops towards the tails. 
It is clear that for a perfect convergence, one needs to achieve $n\rightarrow\infty$ and 
$N\rightarrow\infty$ limits. 
In order to further strengthen these results, we also plot in Fig.~3b the $q$-logarithm 
[defined to be $\log_q(x)=(x^{1-q}-1)/(1-q)$] of the case which is the closest to 
the critical point that we can numerically obtain. 
A satisfactory agreement between the data and $q$-Gaussian can easily be appreciated 
for the case with the largest $n$ (better convergence in the tails could have been achieved 
if cases that are closer to the critical point would have been numerically attainable).
In Fig.~4 we also give the same demonstration of Fig.~3 for the closest case to the critical point 
where we have used $\Omega_{\infty}=\Omega_{c}=0.606661063470$ with three different $N$ values 
which are smaller than $N^*$ since the value of $N^*$ is beyond the present computational 
limits. 
As expected, the limit distribution would converge to the appropriate $q$-Gaussian for the 
{\it entire region} whenever $N=N^*$. Unfortunately, for the present case, since we are very 
close to the critical point, the appropriate $N^*$ value is far beyond our 
computational limits. Nevertheless, as $N$ values increase, the convergence to the   
appropriate $q$-Gaussian (which is with $q=0.90$), starting from the central part, can easily be seen. 
For the largest $N$ value that we can numerically attain, almost the entire 
region (the central part and the tails) is well approached to the $q$-Gaussian 
distribution.

The observed behavior obtained here for the sine-circle map appears to be exactly similar to that 
of the logistic map in the sense that the numerical results agree better and better with 
the appropriate $q$-Gaussian as the critical point is approached although the route to chaos 
here is via quasi-periodicity whereas it is via period doublings in the logistic map. 
The only difference is that the $q$ values obtained here are below unity, 
whereas for the logistic map they are always above unity.

\begin{figure}
\includegraphics*[height=6cm]{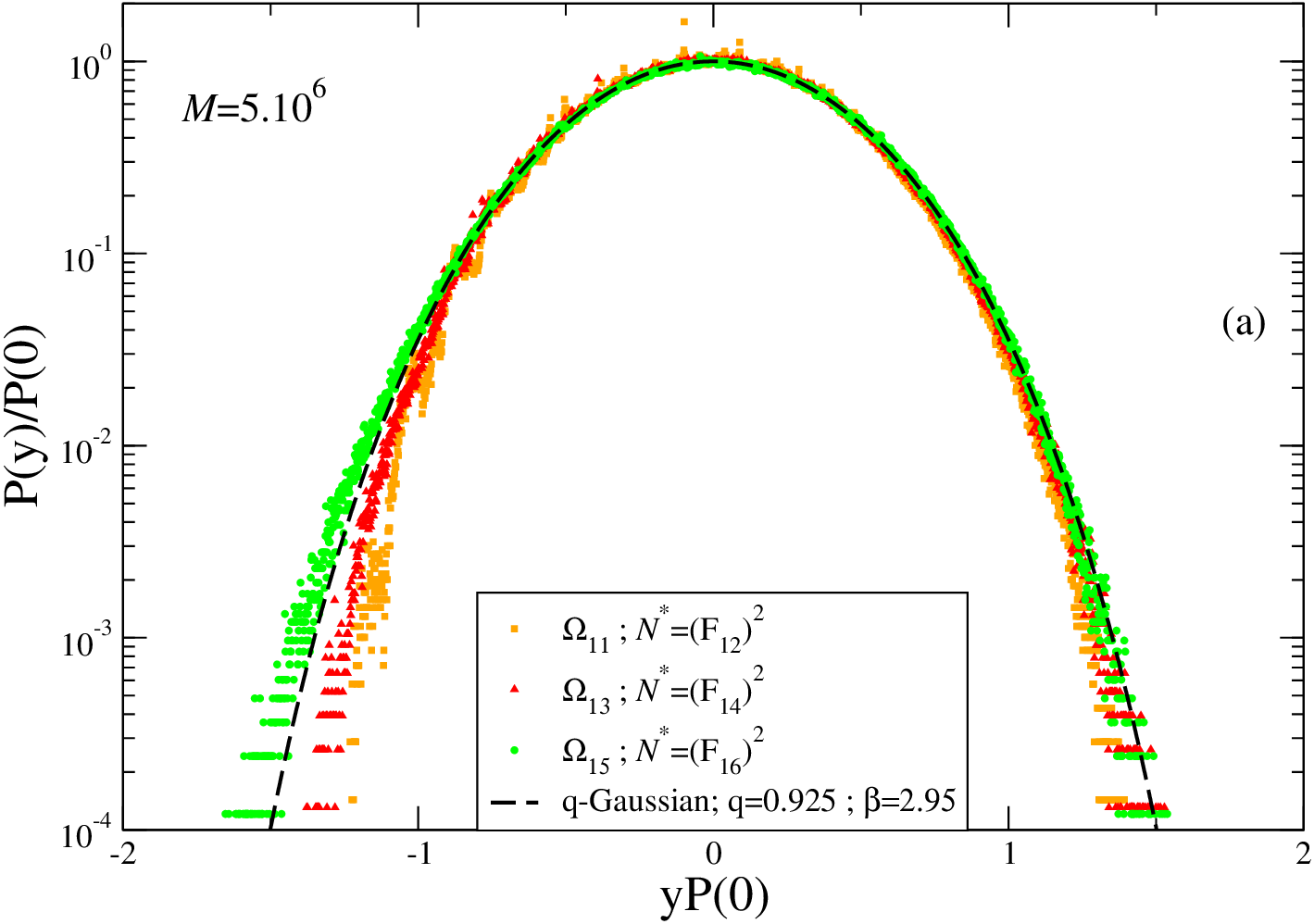}
\includegraphics*[height=6cm]{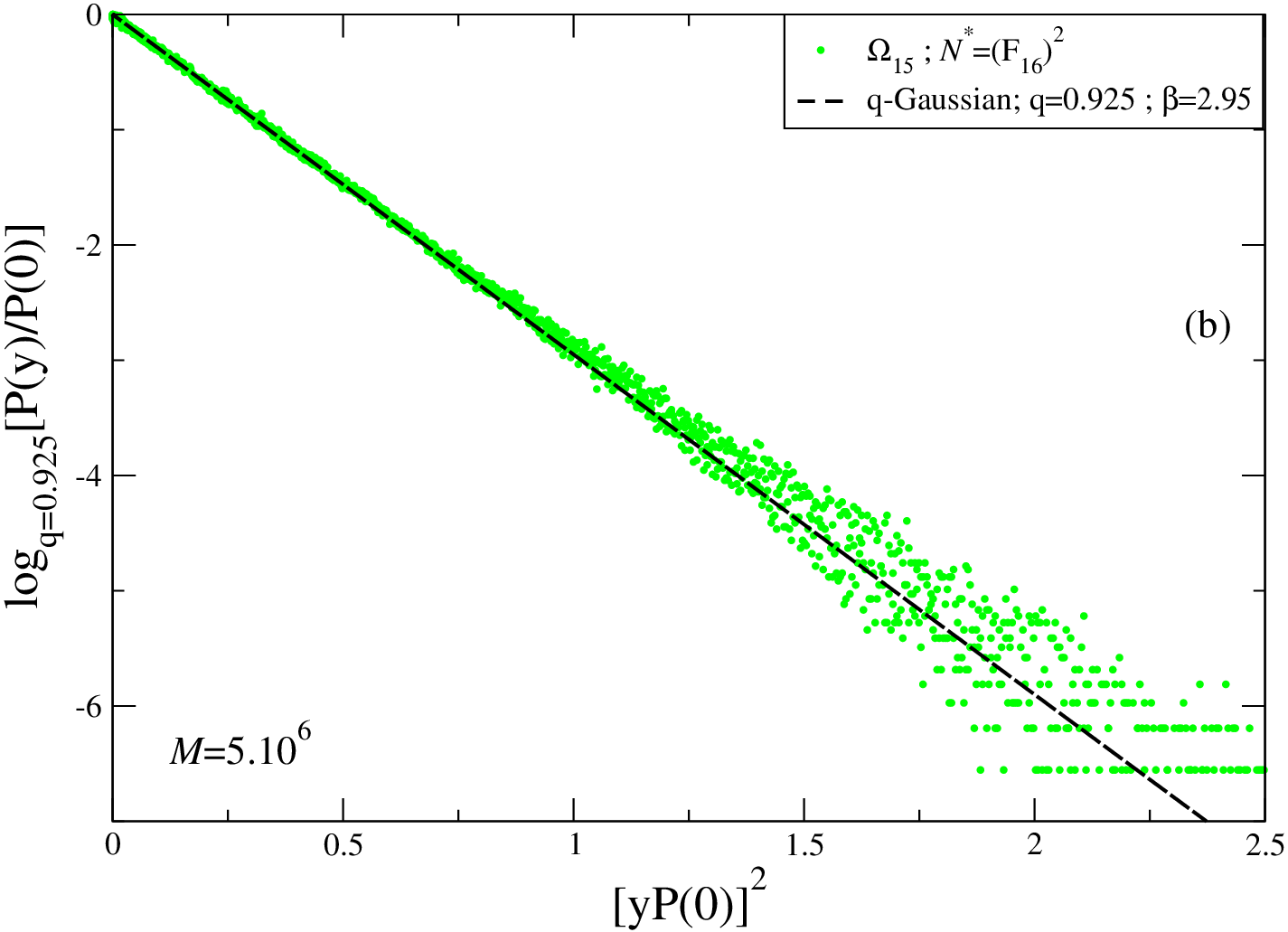}
\caption{\label{fig:Fig3}(Color online) (a)~Probability distributions of the cases 
$\Omega_{11}$, $\Omega_{13}$ and $\Omega_{15}$ with appropriate $N^*$ values. 
(b)~$q$-logarithmic plot of the same data for the case $\Omega_{15}$ which is the closest 
to the critical point that we can numerically obtain in our simulations.}
\end{figure}

\begin{figure}
\includegraphics*[height=6cm]{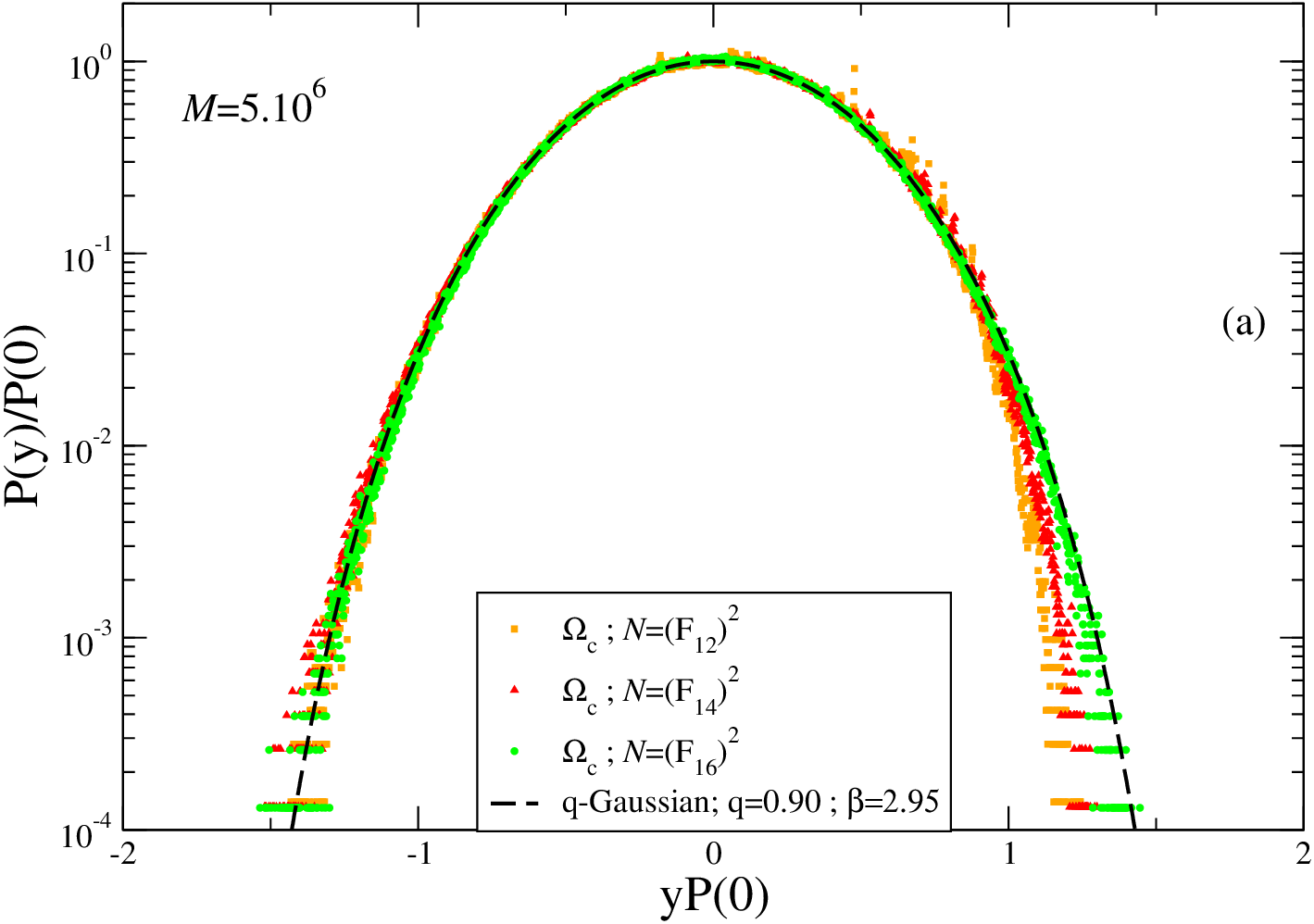}
\includegraphics*[height=6cm]{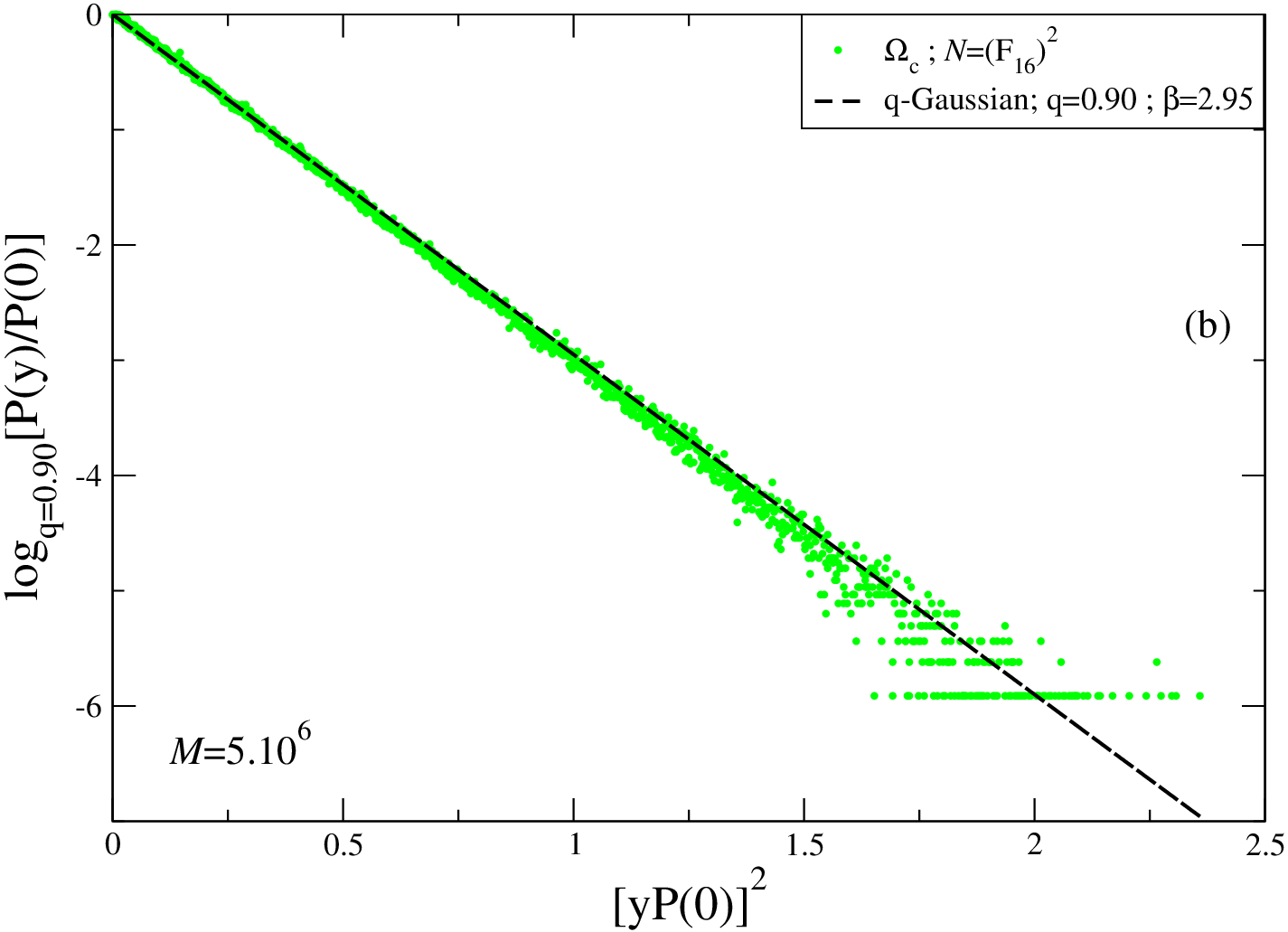}
\caption{\label{fig:Fig4}(Color online) (a)~Probability distributions of the case 
$\Omega_{c}$ with increasing $N$ values with $N<N^*$ (for this value of $\Omega$ the appropriate 
$N^*$ value is far beyond the present computational limits). 
(b)~$q$-logarithmic plot of the same data for the largest possible $N$ that 
we can numerically obtain in our simulations. }
\end{figure}

\section{Conclusions}
Main results obtained in this work can be summarized as follows: 
(i)~The sine-circle map, which belongs to the same universality class of the 
forced Rayleigh-Benard convection, is investigated and it is shown that 
its appropriate probability distribution approaching the quasi-periodic edge of chaos 
appears to be $q$-Gaussian. 
Up to now, in the literature, the only chaos route investigated within this 
procedure was the period doubling route where the $q$ values seem to be always 
larger than unity. 
Since the quasi-periodic route to chaos is different from the period-doubling route to chaos, 
our analysis performed here can be thought as a step forward towards the inclusion of 
another route to chaos (with $q<1$) to the phenomenon observed so far for the period-doubling 
route to chaos (with $q>1$). In this sense, our results broaden the applicability region of 
the arguments argued in \cite{tirbectsal,tsatir} and provide further evidence to the 
proposition that, for certain classes of strongly correlated random
variables, the rescaled sum of the iterates of such dynamical systems approaches a $q$-Gaussian. 

(ii)~Relationship between quasi-periodicity degree of trajectories 
and complete shape of appropriate probability distribution is investigated.
It is seen that the probability distribution converges to a $q$-Gaussian 
as irrationality degree of frequency-ratio parameter increases, which also means 
that the winding number approaches the golden mean value (in other words, 
this is to say as the critical point is approached).  
\\

\section*{Acknowlegments}
This work has been supported by TUBITAK (Turkish Agency) under the Research Project number 104T148 
and by Ege University under the Research Project number 2009FEN077.



\begin{thebibliography}{99}
\bibitem{cramer} H. Cramer,  The Annals of Probability {\bf 4}, 509 (1976).

\bibitem{liapunov} A. Liapunov, Mem. Acad. Sci. St. Petersburg {\bf 12}, 1 (1901).

\bibitem{lindeberg} J.W. Lindeberg, Math. Z. {\bf 15}, 211 (1922). 

\bibitem{billingsley} P. Billingsley, {\it Convergence of Probability Measures}
(Wiley, New York, 1968)

\bibitem{kac} M. Kac, Ann. Math. {\bf 47}, 33 (1946).


\bibitem{beck} C. Beck, Physica A {\bf 169}, 324 (1990).

\bibitem{roepstroff} C. Beck and G. Roepstorff, Physica A {\bf 145}, 1 (1987).

\bibitem{mackey} M.C. Mackey and M. Tyran-Kaminska, Phys. Rep. {\bf 422}, 167 (2006).

\bibitem{cortines} A.A.G. Cortines and R. Riera, Physica A {\bf 377}, 181(2007).

\bibitem{caruso} F. Caruso, A. Pluchino, V. Latora, S. Vinciguerra, and A. Rapisarda,
Phys. Rev. E {\bf 75}, 055101(R) (2007).

\bibitem{pluchino} A. Pluchino, A. Rapisarda and C. Tsallis, Europhys. Lett. {\bf 80}, 26002 (2007).


\bibitem{umarov} S. Umarov, C. Tsallis, and S. Steinberg, Milan J.
Math. {\bf 76} (2008) 307; 
S. Umarov, C. Tsallis, M. Gell-Mann and S. Steinberg, J. Math. Phys. {\bf 51}, 033502 (2010); 
K. P. Nelson and S. Umarov, Physica A  {\bf 389}, (2010) 2157; 
S. Umarov and C. Tsallis, in {\it Complexity, Metastability and Nonextensivity},
eds. S. Abe, H.J. Herrmann, P. Quarati, A. Rapisarda and C. Tsallis,
American Institute of Physics Conference Proceedings {\bf 965}, 34 (New York, 2007); 
S. Umarov and C. Tsallis, Phys. Lett. A {\bf 372}, 4874 (2008).

\bibitem{tsallis} C. Tsallis and S.M.D. Queiros, in
{\it Complexity, Metastability and Nonextensivity},
eds. S. Abe, H.J. Herrmann, P. Quarati, A. Rapisarda and C. Tsallis,
American Institute of Physics Conference Proceedings {\bf 965}, 8 (New York, 2007).

\bibitem{queiros} S.M.D. Queiros and C. Tsallis, in
{\it Complexity, Metastability and Nonextensivity},
eds. S. Abe, H.J. Herrmann, P. Quarati, A. Rapisarda and C. Tsallis,
American Institute of Physics Conference Proceedings {\bf 965}, 21 (New York, 2007).

\bibitem{vignat} C. Vignat and A. Plastino, J. Phys. A {\bf 40}, F969-F978 (2007).

\bibitem{tccmp} C. Tsallis, J. Stat. Phys. {\bf 52}, 479-487 (1988);
E.M.F. Curado and C. Tsallis, J. Phys. A {\bf 24}, L69 (1991);
Corrigenda: {\bf 24}, 3187 (1991)  and {\bf 25}, 1019 (1992);
C. Tsallis, R.S. Mendes and A.R. Plastino, Physica A {\bf 261}, 534 (1998).

\bibitem{tsallisbook} C. Tsallis, {\it Introduction to Nonextensive Statistical Mechanics - 
Approaching a Complex World} (Springer, New York, 2009).

\bibitem{tibet} U. Tirnakli, C. Beck and C. Tsallis, Phys. Rev. E {\bf 75}, 040106(R) (2007).

\bibitem{tirbectsal} U. Tirnakli, C. Tsallis and C. Beck, Phys. Rev. E {\bf 79}, 056209(R) (2009).

\bibitem{tsatir} C. Tsallis and U. Tirnakli, J. Phys.: Conf. Ser. {\bf 201} (2010) 012001.

\bibitem{robledo1} M.A. Fuentes and A. Robledo, J. Stat. Mech., P01001 (2010). 


\bibitem{tsalruiz} G. Ruiz and C. Tsallis, Eur. Phys. J. B, 00054-2 (2009). 

\bibitem{andrea} G. Miritello, A. Pluchino and A. Rapisarda, Physica A {\bf 388}, 4818 (2009). 

\bibitem{silvio} S.M.D. Queiros, Phys. Lett. A, {\bf 373}, 1514 (2009). 


\bibitem{hilborn} R.C. Hilborn, {\it Chaos and Nonlinear Dynamics} 
(Oxford University Press, New York, 1994), p. 390. 

\bibitem{schuster} H.G. Schuster, {\it Deterministic Chaos: An Introduction} 
(VCH, Weinheim, 1988). 


\bibitem{jensen} M.H. Jensen, L.P. Kadanoff, A. Libchaber, I. Procaccia and J. Stavans, 
Phys. Rev. Lett. {\bf 55}, 2798 (1985).

\bibitem{circle1} U. Tirnakli, C. Tsallis and M.L. Lyra, Eur. Phys. J. B {\bf 11}, 309 (1999). 

\bibitem{robledo} H. Hernandez-Saldana and A. Robledo, Physica A {\bf 370}, 286 (2006).



\end{thebibliography}
\end{document}